\documentclass[prl,twocolumn,showpacs]{revtex4}
\usepackage{dcolumn}
\usepackage{graphicx}
\usepackage{psfig}
\usepackage{amsmath,epsf}
\usepackage{amssymb}
\begin{document}

\title{Anomalous Ferromagnetism of Monatomic Co Wire at the Pt(111)
Surface Step Edge}
\author{Alexander B. Shick,$^1$ Franti\u{s}ek M\'aca,$^1$ and
Peter M. Oppeneer$^2$}
\affiliation{$^1$Institute of Physics, ASCR, Na Slovance 2, CZ-182 21
Prague 8, Czech Republic\\
$^2$Leibniz-Institute of Solid State and Materials
Research, P.O. Box 270016, D-01171 Dresden, Germany}
\begin{abstract}
A first-principles investigation of the anomalous
ferromagnetism of a quasi-one-dimensional
Co chain at the Pt(111) step edge is reported.
Our calculations show that
the symmetry breaking at the step leads to an easy magnetization
axis at an odd angle of $\sim$20$^{\circ}$ {\em towards} the Pt step,
in agreement
with experiment [P. Gambardella {\em et al.}, {\em Nature} {\bf 416}, 301
(2002)]. Also, the Co spin and orbital moments become noncollinear,
even in the case of a collinear ferromagnetic spin arrangement.
A significant enhancement of the Co orbital magnetic moment is
achieved when modest electron correlations are treated within
LSDA+$U$ calculations.
\end{abstract}
\date{\today}
\maketitle

Exploring magnetism in the one-dimensional (1D)
limit has been a great challenge for many years.
Only recently, Gambardella {\it et al}.~\cite{gambardella02}
succeeded to observe ferromagnetism
of monatomic Co wires decorating the Pt(997) surface step edge.
By exploiting the element-selectivity of the
x-ray magnetic circular dichroism (XMCD),
the existence of long-range ferromagnetic order on Co was
demonstrated below 15~K
\cite{gambardella02,gambardella03}.
Although theoretically the Mermin-Wagner theorem \cite{mermin66}
forbids long-range 1D ferromagnetic order at non-zero temperatures,
ferromagnetism in 1D can be stabilized by a large magnetic
anisotropy energy, which creates barriers effectively blocking
thermal fluctuations. The significance of such blocking
mechanism was recognized earlier for the occurrence of
long-range magnetic order in 2D systems \cite{hauschild98,schneider00}.

The experiments of Gambardella {\it et al}. revealed novel
magnetic properties of monatomic Co wires at Pt step edges. An
unexpected magnetocrystalline anisotropy was observed: the easy
magnetization axis was directed along a peculiar angle of
+43$^{\circ}$ towards the Pt step edge and normal to the Co chain.
The magnetocrystalline anisotropy energy (MAE) was estimated to be
substantial, of the order of 2~meV/Co atom \cite{gambardella02}.
In addition, a considerable enhancement of the Co orbital magnetic
moment $M_L \approx 0.7$~$\mu_B$---as compared to the bulk Co
$M_L$ value of 0.14 $\mu_B$---was deduced from XMCD experiments.

In this paper we report a first-principles investigation of the
anomalous ferromagnetism of a monatomic Co wire at the Pt(111)
surface step edge, using state-of-the-art electronic structure
calculations. We focus on the intriguing features of the quasi-1D
Co wire, i.e., the easy axis rotated away from the (111) surface
normal, the enhanced Co orbital moment and huge estimated MAE. The
key outcomes of our study are {\it (i)} the {\em ab initio}
calculation of an easy axis at an odd angle rotated towards the Pt
step edge and {\it (ii)} the prediction of an intrinsic
noncollinearity between spin and orbital magnetic moments of both
the ferromagnetic Co wire and Pt substrate. The origin of this
novel magnetic behavior, which is to our knowledge not present in
known 2D and 3D itinerant ferromagnets is explained to be a
consequence of the magnetic symmetry lowering at the surface step
edge \cite{remark}. 
Our calculations furthermore yield a MAE of the order of 4
meV/Co atom, and---using the LSDA+$U$ approach---a Co orbital moment
$M_L = 0.45\,\mu_B$.

Previously, several computational studies of the magnetic
properties of adatoms, clusters, and monatomic chains on surfaces
were reported (see, e.g.,
\cite{dorantes98,cabria02,spisak02,lazarovits03,hong03}). The
calculations predict in general an enhanced MAE closely related to
the reduced dimensionality and
enhancement of the orbital moment. However, in all of these
studies only transition-metal wires or adatoms on {\it flat}
surfaces are investigated, i.e., geometries that are essentially
different from the metal wire at a step edge. For wires, adatoms
or clusters \cite{cabria02} at flat surfaces the easy axis is
either normal to the surface or in-line for some wires
\cite{hong03}. So far only one {\it ab initio} study of Co at a
Pt step edge was reported, in which the XMCD spectrum was computed
\cite{komelj02}, but the magnetic anisotropy was not considered.
Our study focuses on the unprecedented magnetic anisotropy
properties observed for the quasi-1D Co chain.

\begin{figure}[t]
\includegraphics[width=8cm,height=9.5cm]{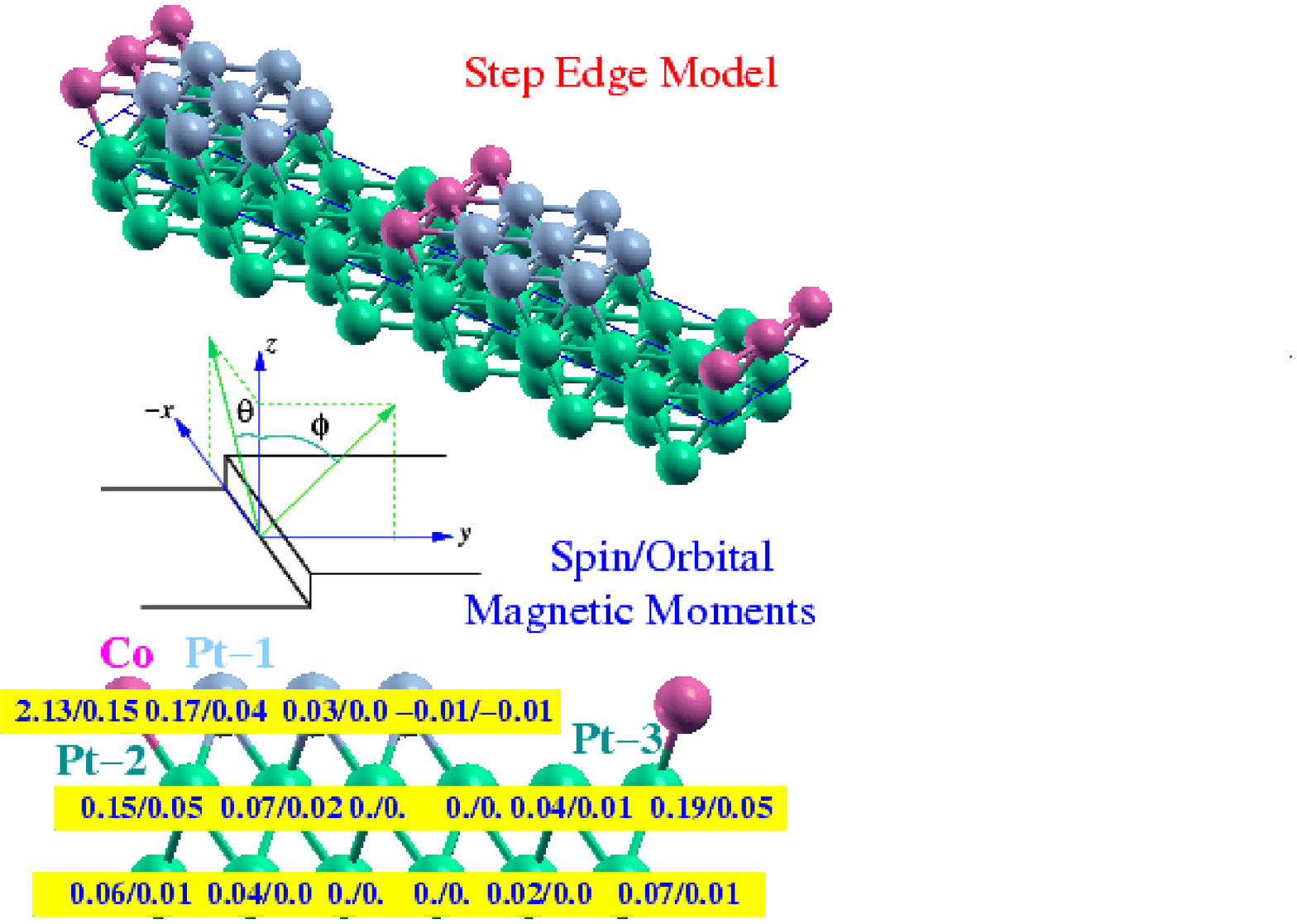}
\caption{Top: schematic crystal structure of model II,
used to represent the Co chain at the Pt(111)
surface step edge \cite{kokalj99}.
Middle: Definition of the angles $\theta$, $\phi$, and
coordinate axes. The ${x}$ axis is chosen
parallel to $[1 \bar{1} 0]$ (along the Co wire), the ${y}$ axis to
$[1 \bar{2} 1]$ (normal to the wire), and the surface normal
${z}$ is chosen parallel to $[111]$.
Bottom: profile sketch with calculated $M_S^{z}/M_L^{z}$ values
specified. } \label{fig1}
\end{figure}

{\em Methodology.} We performed supercell calculations to model
the Co chain at the Pt(111) surface step edge.
Supercells of various sizes were investigated. We shall discuss
here particularly two supercells: one of small size, a toy model,
which we name model I, and a large, realistic supercell, model II
(see Fig.~1). Model I consists of one subsurface
Pt layer built of 4 rows of Pt and one surface layer containing
one row of Co atoms, two rows of Pt, as well as one empty row to
model the step edge. Model II consists of a sub-subsurface and a
subsurface Pt layer built of 6 rows of Pt atoms, while the surface
step is modeled by 3 rows of Pt, one Co row, and two rows of
empty Pt sites. In both supercells the vacuum is modeled by the
equivalent of two empty Pt layers. All interatomic distances are
adopted to be those of pure Pt. We note that while model II
approaches the maximally treatable supercell size for
full-potential, relativistic calculations of
the MAE, the proportions of the experimental Co chain at the
Pt(997) step edge are  still larger, consisting of an 8 Pt rows wide
terrace at the Pt step \cite{gambardella03}.

The first-principles calculations were performed using the
relativistic full-potential linearized-augmented-plane-wave
(FP-LAPW) method, in which the spin-orbit coupling is included in
a self-consistent second-variational procedure \cite{shick97}. For
most of the calculations the conventional
(von Barth-Hedin) local spin-density
approximation (LSDA) is adopted, which is expected to be valid for
itinerant metallic systems. In order to capture better the
electron correlations expected for the Co $3d$ electrons in the
reduced dimension also the LSDA+$U$ approach, in the
implementation of Ref.~\onlinecite{shick01} has been applied.  For
further details of the calculations, see
Ref.~\onlinecite{details}.

{\it First-principles results.} We first applied the conventional
LSDA approach using the FP-LAPW method. To start
with, the spin magnetization axis was chosen to be fixed either
along the $x$, $y$, or $z$ axis. The essential computed spin
($\vec{M}_S$) and orbital ($\vec{M}_L$) moments are given in Table
\ref{table1}. Table \ref{table1} reveals that the $\vec{M}_S$ and
$\vec{M}_L$ on Co and Pt are {\it noncollinear} for a spin moment
fixed along the $y$ or $z$ axis, but {\it collinear} when
$\vec{M}_S$ is along the $x$ axis. Noncollinearity of $\vec{M}_S$
and $\vec{M}_L$ has been predicted previously \cite{sandratskii98}
for materials exhibiting a noncollinear spin magnetic structure,
but this is to our knowledge the first observation of such
noncollinearity for a collinear, ferromagnetic spin configuration.
To understand the noncollinearity of $\vec{M}_S$ and $\vec{M}_L$
it is instructive to consider the magnetic symmetry.
The symmetry operations which preserve the crystal symmetry are
the identity $E$ and the mirror operation $\sigma_{x}$ with
respect to the $yz$ plane (see Fig.~1). Considering now the
magnetic symmetry
operations, which are---for a total magnetic moment in the
$yz$ plane---$E$ and $\sigma_{x}R$, with $R$ the time inversion,
we observe that these symmetry conditions impose $M^x$=0, but
$M^y,\,M^z\ne$0 without any restriction. Therefore there is no
particular symmetry imposed direction in the $yz$ plane which
would force spin and orbital moment to be parallel. In other
words, the magnetic symmetry in the $yz$ plane is the same for all
magnetization directions.
Along the wire the situation is different: the magnetic symmetry
operations $E$ and $\sigma_{x}$, which conserve $M^x$, force
$M^y$=$M^z$=0 and consequently, we must have $\vec{M}_L \parallel
\vec{M}_S$ for a magnetization along the wire.

\begin{table}[b]
\caption{Spin and orbital magnetic moments in $\mu_B$, calculated
for the monatomic Co wire at the Pt(111) step edge, using the supercell
models I and II. }
\begin{tabular}{ccccccccc}
\hline
       & & \multicolumn{3}{c}{$\vec{M}_S$} & \multicolumn{1}{c}{~~}
       & \multicolumn{3}{c}{$\vec{M}_L$}\\
\cline{3-5}\cline{7-9}
Model I, axis &  & $x$ & $y$ & $z$ &   & $x$   & $y$  & $z$  \\
\hline\\[-0.2cm]
$\vec{M}_S \parallel  x$ axis &  & 2.129 & 0 & 0 & & .084 & 0    & 0  \\
$\vec{M}_S \parallel  y$ axis &  & 0 & 2.128 & 0 & & 0    &0.065 & 0.032  \\
$\vec{M}_S \parallel  z$ axis &  & 0 &   0 & 2.127& & 0     &0.009 & 0.155  \\
\hline
\\[-0.2cm]
Model II, atom\footnote{$\vec{M}_S$ parallel $z$ axis} & & & & & & & &\\
\hline
Co & & 0 &   0 & 2.127  && 0     & 0.011 & 0.149  \\
Pt-1 &  & 0 &   0 & 0.168 & & 0     & 0.005 & 0.044  \\
Pt-2 &  & 0 &   0 & 0.146 & & 0     & 0.003& 0.046  \\
Pt-3 &  & 0 &   0 & 0.194 & & 0     & 0.005 & 0.052  \\
\hline
\end{tabular}
\label{table1}
\end{table}

From Table \ref{table1} we further observe that both the Co
spin and orbital moment are considerably enhanced with respect to
the values for bulk hcp Co \cite{trygg95}, as expected
for a dimensionality reduction leading to a more atomic-like
configuration. The calculated $M_S$ and $M_L$ of Co agree well
with those of Ref.~\onlinecite{komelj02}, where however only
$z$ axis collinear components of $M_S$ and $M_L$ were
considered. The Co $M_S$ does not change when the
supercell is enlarged from model I to model II, but small changes
in the orbital moments exist. Also, there is a sizable
magnetization induced on the nearest neighbor Pt atoms, which
is decreasing rapidly for the Pt atoms farther away (see Fig.~1). The
size of supercell model II appears thus sufficient to separate the
magnetic Co wires and to ensure the Pt magnetization decrease away
from the step edge.

\begin{figure}[t]
\label{fig2}
\includegraphics[width=7cm,height=6cm]{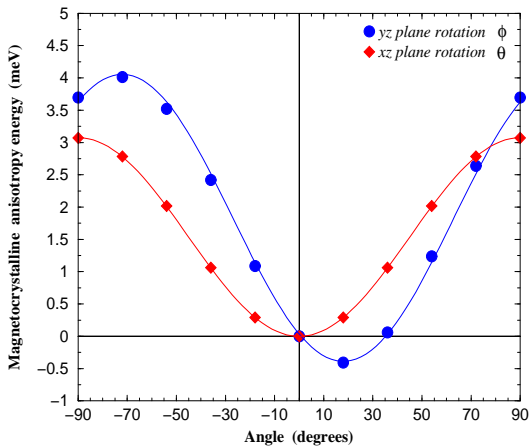}
\caption{The MAE calculated for model II
as a function of the angles $\theta$ and $\phi$ of the
spin moment. The solid lines are fits to the
points and are given by MAE $=4.04 -4.44\cos^2(\phi -18^{\circ})$
in the $yz$ plane and $3.08 - 3.08\cos^2 \theta $ in the $xz$ plane.
}
\end{figure}

Next, we turn to the salient aspect of our investigation, the MAE
calculations. We used the so-called ``magnetic force theorem" to
compute the MAE: starting from self-consistent charge and spin
densities calculated for the spin moment aligned along the ${z}$
axis, the $M_S$ is rotated over angles $\theta$ or $\phi$ (see
Fig.~1) and a single energy band calculation is performed for the
new orientation of $M_S$.
The MAE, which is defined as a directional total-energy difference,
is computed from the change in one-electron energies $E$
due to the $M_S$ rotation, i.e.,  MAE $= E(\theta, \phi) -
E(\theta=0, \phi=0)$ \cite{details2}.
The calculated MAE is shown in Fig.~2 for supercell II.

For a rotation of $M_S$ over an angle $\theta$ in the $xz$ plane
the MAE dependence on $\theta$ is symmetric, reflecting the mirror symmetry
$\sigma_x$, with the easy axis pointing along the
${z}$ direction and the hard axis directed along the Co wire.
A MAE difference between the hard and easy axes of $\sim$3
meV/Co ($\sim$1 meV/Co for model I) is calculated, exceeding
by an order of magnitude the dipolar shape anisotropy \cite{balazs}.

For a rotation of $M_S$ over an angle $\phi$ in the $yz$ plane,
we obtain a peculiar {\em asymmetric} dependence of the MAE on $\phi$
(see Fig.~2), 
reflecting the absence of any particular symmetry imposed direction
in the $yz$ plane. The computed easy axis is rotated away from the
$z$ axis by $18^{\circ}$ towards the Pt step edge, in
semi-quantitative agreement with the experimentally observed
anomalous $43^{\circ}$ easy axis \cite{gambardella02}. The
calculated direction of the hard axis of $-72^{\circ}$ corresponds
reasonably with the experimental value of $\approx -50^{\circ}$
also. A notable difference appears for MAE calculations adopting
model I: while the calculated MAE($\phi, \theta=0$) curve is
asymmetric as well, the minimum occurs at $\approx -50^{\circ}$
(not shown),
thus oriented outwards from the Pt step. In contrast to
supercell model II the toy model I is not valid even
qualitatively for the MAE. Increasing the supercell size from
model I to II improves the MAE, therefore an even better
agreement with experiment can be expected for even larger
supercells.
The MAE difference between the hard and easy axes is $\approx$4.45
meV/Co for model II ($\approx$1.6 meV/Co for model I), which is of the
same magnitude as the experimentally estimated MAE of $\sim$2
meV/Co at $T=45$~K \cite{gambardella02}. We expect a definitely
higher experimental MAE and thus an even better agreement with model II
for $T=0$~K.
We note that previous studies showed the conventional LSDA theory to
be quite successful for describing the uniaxial MAE of hcp Co and CoPt
bulk alloy \cite{shick03,oppeneer98}. Here we demonstrate that the LSDA
also provides a reasonable explanation of the MAE of the Co wire
at the Pt step edge.

Although the LSDA works well for MAE calculations, it does not
give large enough values for the orbital moment \cite{trygg95}.
For example, the LSDA calculated $M_L$ of hcp Co is with 0.08
$\mu_B$ only half the experimental value of 0.14 $\mu_B$. The
situation is even worse for the Co wire. Comparing the LSDA
calculated $M_L$ $\sim$0.15 $\mu_B$ (see Table I) with the
experimental  $M_L$ of $0.68\pm0.05$ $\mu_B$ \cite{gambardella02}
indicates that the LSDA value is too small by a factor of 4.5!
Recently, the orbital-polarization correction (LSDA+OP) was
applied to a Co wire on Pt, leading to a Co $M_L$ of 0.92 $\mu_B$
\cite{komelj02}, but this value overshoots the experimental data.

To improve the $M_L$ one needs to account for
electron-correlation effects beyond the conventional LSDA, which
is currently a challenging problem of {\em ab initio}
relativistic energy-band theory. To estimate this effect, we use here the
{\em semi-model} but physically transparent LSDA+$U$ method, which
was shown to correct the Co $M_L$ of both hcp Co and CoPt alloy with a
{\it single} choice of the Coulomb $U$ ($=1.7$ eV) and exchange
$J$ ($=0.91$ eV) parameters \cite{shick03}.
Using the same $U$ and $J$, we compute $M_L^z$= 0.45 $\mu_B$ for
supercell II when $M_S$ is along the $z$ axis ($M_L^z=$ 0.32
$\mu_B$ for model I). These values are still smaller than the
experiment but we expect a larger $M_L$ with a further increase of
the supercell, due to a related Co $d$-states localization. We
could of course obtain better agreement with experiment by another
choice of $U$ and $J$, but we prefer to use the ``universal''
values found in Ref.~\onlinecite{shick03}, treating thus the $U$
of metallic Co as a transferable, atom specific quantity.
The Co spin moment hardly changes, from 2.13 to 2.18 $\mu_B$ when
the $U$ is included. We note, that in the LSDA+$U$ method the MAE
can be computed only from total energies, as it is incompatible
with the force theorem. It will require highly accurate self-consistent
calculations for the $M_S$ rotated over different angle $\theta$ and $\phi$,
something which is numerically not a practicable approach.                   

\begin{figure}[t]
\includegraphics[width=6.5cm,height=9cm]{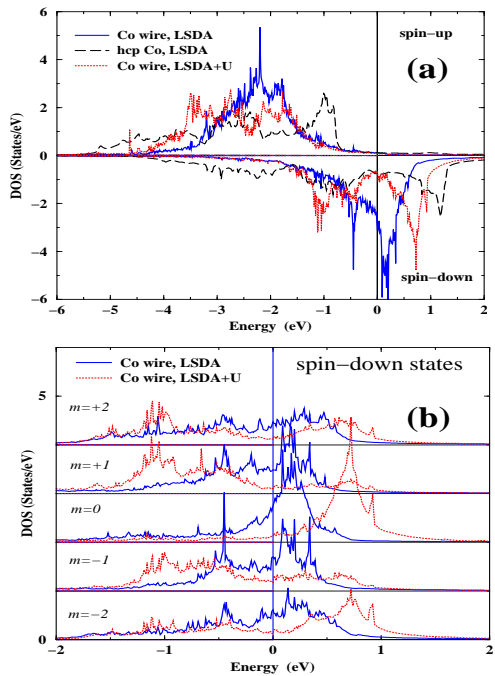}
\caption{(a) The spin-resolved $3d$ DOS, computed with the LSDA and LSDA+$U$
for model II, and compared with the LSDA DOS of hcp bulk Co.
(b) The orbitally-resolved, spin-down DOS calculated
with the LSDA and LSDA+$U$ methods.}
\end{figure}

To understand how the enlargement of the Co $M_L$ in the LSDA+$U$ approach
comes about we consider the spin and orbitally resolved
$3d$ densities of states (DOS) of model II, which are shown in Fig.~3.
The spin-resolved LSDA DOS reveals a substantial narrowing of the band
width from $\sim$6 eV for hcp Co to $\sim$4 eV for the Co wire as well
as a moderate increase of the spin-splitting (Fig.~3a), as is expected
for the reduced Co coordination. When the $U$ is included the $3d$ DOS
broadens somewhat and significant changes in the spin-resolved DOS occur.
Since the spin-up Co $d$-band is fully occupied, only changes of
the spin-down band are essential for the $M_L$ enhancement.
The spin-down $m$-resolved $3d$ DOS of the Co wire
is shown in Fig.~3b. A major change in the LSDA+$U$ DOS
appears as an upward shift within the $|\!\! \downarrow; m=0\rangle$ DOS,
which, however, does not contribute to $M_L$.
The major contribution to the increase of $M_L$ originates from an upward
shift within the $|\! \!\downarrow ; m=-2 \rangle$ and a downward shift of
the  $|\!\! \downarrow ; m=+2 \rangle$ DOS. Also, downward shifts of
the $|\!\! \downarrow ; m=\pm 1 \rangle$ states take place, but these
contribute only secondarily to the $M_L$ change.
Thus, we conclude that the $M_L$ enhancement
with the Coulomb $U$ is brought about by
modifications of the in-plane spin-down ${x^2-y^2}$ and
${xy}$ orbital densities and much less affected by changes in the
out-of-plane ${xz,yz}$, and ${3z^2-r^2}$ orbital densities.
The in-plane orbitals are affected most by the
missing Pt atoms at one edge side and thus most liable to localize.

In conclusion, employing first-principles calculations we have
provided a microscopic picture of the anomalous magnetocrystalline
anisotropy of a quasi-1D Co chain at the
Pt(111) step edge. The essential symmetry breaking at the step edge
leads to noncollinear spin and orbital moments as well as to
an easy magnetization axis oriented
at a peculiar angle towards the Pt step edge. LSDA theory is found
to provide a rather good explanation of the magnetocrystalline anisotropy,
yet a considerable improvement of the Co orbital moment is obtained
with LSDA+$U$ calculations.
\\
\indent
We gratefully acknowledge discussions with P. Gambardella, P.
Nov\'ak, H. Eschrig, and W.E. Pickett. This work was supported by
the State of Saxony and the Grant Agency of the ASCR Grant
A1010214.


\begin{thebibliography}{99}
\bibitem{details2} We used 196 and 192 special $k$-points
for model I and II, respectively, in a quasi-2D BZ with $k_z=0$.
\end{thebibliography}
\end{document}